# POLYANA - A tool for the calculation of molecular radial distribution functions based on Molecular Dynamics trajectories

Christos Dimitroulis[1], Theophanes Raptis[1], Vasilios Raptis[2]

[1] Computational Applications Group, Division of Applied Technologies, National Centre for Science and Research "Demokritos", 153 10, Aghia Paraskevi, Athens, Greece

[2] Institute of Nanoscience and Nanotechnology, National Centre for Science and Research "Demokritos", 153 10, Aghia Paraskevi, Athens, Greece.

Corresponding author: Vasilios Raptis (e-mail: v.raptis@inn.demokritos.gr)

**Abstract**:

We present an application for the calculation of radial distribution functions for molecular centres of mass, based on trajectories generated by molecular simulation methods (Molecular Dynamics, Monte Carlo). When designing this application, the emphasis was placed on ease of use as well as ease of further development. In its current version, the program can read trajectories generated by the well-known DL_POLY package, but it can be easily extended to treat other formats. It is also very easy to 'hack' the program so it can compute intermolecular radial distribution functions for groups of interaction sites rather than whole molecules.

## 1. Introduction

Molecular Dynamics and Monte Carlo atomistic simulations are an integral part of nowadays research in the areas of molecular liquids, complex molecular systems, materials etc. Among others, one can look at the microscopic structure of such systems, usually quantified in the form of pair distribution functions and, in the case of homogeneous systems, radial distribution functions (RDF) for each kind of interaction site pairs defined on the basis of types of sites present in the system. In a similar fashion, one is often interested in the computation of RDFs for whole molecules rather than individual interaction sites, especially in the case of small species. Molecular RDFs, $g(r,\omega)$ are, in general, functions of both the molecular pair distance, $r$, and mutual orientation, $\omega$. However, their simpler version, $g(r)$, in which orientational degrees of freedom are averaged out and only pair distance is taken into account are by no means less important. Structure generation and refinement and development of pairwise-additive force fields by means of Reverse Monte Carlo [1-6] or the variants of Boltzmann inversion technique [7-12] are examples worth mentioning, among others.

For instance, when developing a coarse-grain model, centres of mass are required instead of individual site coordinates, but orientational degrees of freedom are eliminated.

The DL_POLY package [13-15] is a well-known program for the simulation of molecular systems by means of the Molecular Dynamics method. With the end of a DL_POLY run, the RDFs of all interaction site pairs are computed by the program and stored in a file called 'RDFDAT'. However, the program does not compute centre-of-mass RDFs and, to the authors' knowledge, no application that could carry out this task for DL_POLY trajectories, seemed to be available at the time of writing this article. That was one of the motivations for the present work: to help the DL_POLY community by offering a freely available tool to compute molecular RDFs.

The second motivation was the authors' desire to start developing a molecular simulation toolbox of post-processing applications designed with simplicity of usage in mind – ideally, the user manual could be summarised in one sentence: type the program's name in the command prompt and press 'Enter'. In this article, we present POLYANA[1], one such easy to use application that can read DL_POLY output files and compute molecular RDFs for all species present in the simulated system without the users having to define any particular input unless they want to. By properly tweaking the DL_POLY input files, one can also compute RDFs for specific groups of sites rather than whole molecules.

Our last motivation was to start building our toolbox in such a way that it will be easy for us or other interested developers to further expand it by adding new functionality. POLYANA is a self-contained code without external dependencies, written in Fortran2003; the distributed Makefile allows for a Fortran95 version, alhtough freely available compilers like GNU Fortran are updated to comply with the most recent standards. Internally, the POLYANA code is structured as a composition of modules corresponding to physically meaningful entities (atoms, molecules and so on) equipped with member functions to set or compute and get the values of important variables and parameters. These are then used in processing routines collected in a separate module and called as needed by a main 'driver' program.

POLYANA does not need additional input data to run – DL_POLY input and output files serve this purpose. However, if one wants to exert finer control on the program's execution, simple, particularly easy to learn directives can be inserted to this purpose, in the appropriate place of the DL_POLY input files. Currently, DL_POLY comes in two versions: the freely available 'Classic' version [13] and version 4 or DL_POLY_4 [14, 15] that is available to registered academic users.

---

[1] So named after a previous, in-house software of ours aimed at analysing polymer systems, hence called *POLYmer ANAlysis*; the name could be reinterpreted, now, as *dl_POLY ANAlysis*. The package is freely downloadable under the MIT license as gzipped tar file, from: http://cag.dat.demokritos.gr/software/polyana.tar.gz

Both can be executed serially or in parallel but version 4 can scale to a much higher number of processors. The commonest input and output files of the two versions have the same format; POLYANA has been tested against the 'Classic' version but the authors trust that it can handle DL_POLY_4 trajectories as well.

Polyana is suitable for systems of 'small' molecules, i.e. those with sizes that do not exceed half the shortest simulation cell dimension. When such is the case, any molecular topology can be handled by a generic algorithm that lifts the periodic boundary conditions and unfolds molecules to compute their centre of mass. Indeed, it is shown via numerical examples, as explained in the following Section, that the particular algorithm works correctly even for arbitrary topologies, provided the simulation cell is at least twice larger than the molecules' dimensions. Section 2 outlines how the program works internally and Section 3 describes its input and output and examples of usage. Section 4 discusses a particular test case, namely the simulation of a liquid hexane system, to offer illustrative examples of POLYANA usage. Section 5 concludes with a general discussion and future goals.

## 2. Theoretical considerations and POLYANA internals

The concept of pair distribution functions is a very familiar one and it will not be presented here in detail. The interested reader may consult a statistical mechanics textbook such as Hill's [16]. Two particular points have to be taken into account in the calculation of the molecular RDFs. The computation of molecular centres of mass in systems subjected to Periodic Boundary Conditions (PBC); and the computation of intermolecular distances under the same PBCs. The routines for applying or lifting the PBCs have been borrowed from the DL_POLY source code. Currently, POLYANA can handle the cubic, orthorhombic, parallelepiped and slab periodic boundary conditions. Details about them can be found in DL_POLY's user manual.

The two points noted above are treated as follows: First, the molecular centres of mass are computed; then the pair distances among them are calculated and a histogram of distances is updated in order to compute the final RDFs. In order to derive the centre of mass of a given molecule correctly, the molecule has to be re-built by lifting the PBCs so that the correct, 'physical' or 'unfolded' distances among its interaction sites are restored. Usually, one does this by rebuilding the molecule in a bond-by-bond fashion from one end to the other. This is easy to do in the case of chain molecules but it is more difficult in the case of species with complicated connectivities and topologies like the ones involving rings, side chains etc.

However, if the molecule's dimensions are smaller than half the smaller dimension of the simulation cell – and those are with all certainty the cases we are interested in molecular RDFs – there is a more straightforward way to rebuild the molecule without having to take connectivity into account. Indeed, it suffices to compute the pair distances of each interaction site $i = 2, 3, \ldots N_{atmol}$ with respect to a 'base' site $j = i\text{-}1$ and update the unfolded positions accordingly, check whether there are still any 'unphysical' pair distances and repeat until all unfolded pair distances are sorted out, as follows:

(1) FOLDED = FALSE

(2) Multiply all site coordinates $\mathbf{r}_i$ by the inverted simulation cell tensor $C^{-1}$ to compute all reduced coordinates $\mathbf{s}_i = C^{-1}\mathbf{r}_i$.

(3) Calculate the vectors $\mathbf{b} = \mathbf{s}_i - \mathbf{s}_{i-1}$ and $\mathbf{d} = \mathbf{b} - \text{NINT}(\mathbf{b})$

(4) If $| b^2 - d^2 | > $ TOL ( $<< 1$) for every $i$ then

    (4a) FOLDED=TRUE

    (4b) Reduced coordinates are updated as follows: $\mathbf{s}_i = \mathbf{s}_{i-1} + \mathbf{d}$

    (4c) Real coordinates are re-computed by multiplying reduced ones by the simulation tensor $C$.

(5) If FOLDED = TRUE then return to step (1), else exit the loop

The above algorithm has been tested by means of a simple program, included in a separate folder in the distribution. In this particular test, we tried the case of two molecules with random topology repeatedly placed at randomly varying orientations. The centres of mass of the molecules are pegged at a fixed distance and the random orientations are recorded in the form of a DL_POLY style trajectory file. The molecular sizes are defined by the user in the form of a radius of a sphere wherein interaction sites are placed. The resulting RDF graphs always show the expected spike at the given distance corresponding to the centre-of-mass positions. Detailed explanations are provided in the accompanying README file in the source package.

This scheme will only fail in the case of 'large' molecules, i.e. exceeding a critical distance beyond which the minimum image calculation at step (3), above, is not valid [17]. The exact definition of that distance depends on the shape of the simulation cell [18] and, e.g. for a cubic box, it would be half its length. However, in such cases it doesn't make much sense to look at molecular RDFs; these are only of practical interest when many molecules of a given type are dispersed more or less homogeneously in the simulation cell and their number and size relative to the cell's dimensions are such that PBCs do not have any effect on the RDFs within some reasonable distance.

Once the molecule has been rebuilt via the above procedure it is straightforward to compute its centre of mass based on its real unfolded coordinates. Its position then, may be found to lie outside the simulation cell; PBCs are applied to it once again to bring it back in the cell. Given the coordinates of the molecular centres of mass subjected to the system's PBCs, it is easy to compute the molecular distances in the same way as we would for the individual interaction sites. These distances are used to update a histogram of pair occurrences with distance, $h_{\alpha\beta}(r, r+\Delta r)$ where $\alpha$ and $\beta$ are the molecule types and $\Delta r$ is the bin width, and use that histogram to compute the RDFs.

Regarding the RDFs themselves, their computation is divided into two stages: the collection and the averaging stage. The first one takes place when the configurations stored in the trajectory file (called HISTORY in DL_POLY's lingo) are read by POLYANA and the molecular positions are computed as described above. An index, $n = 1+\text{NINT}(r/\Delta r)$ where $r$ is the molecule pair distance, is computed for the respective element of the histograms for pairs $\alpha\beta$ and $\beta\alpha$. Pair distances exceeding a defined cutoff are not taken into account. At the same time, an accumulator, $v$, for the computation of the average cell volume is updated with each configuration – this is useful in the case of the NPT or NσT ensembles. The second operation, averaging, takes place when all configurations have been read. Given a particular molecule type, $\alpha$, the RDF for species $\beta$ around molecules of type $\alpha$ is computed as follows:

First, the histograms and the average volume accumulator are averaged by the number of configurations. The histogram $h_{\alpha\beta}$ is, subsequently, normalised to the number of molecules of type $\alpha$. Then, the numbers of molecules of each type, $a$, are divided by the average volume to convert them to number densities, $\rho_\alpha$. Finally, the elements of the histograms $h_{\alpha\beta}$ are normalised by the shell volumes contained between $(n-1/2)\Delta r$ and $(n+1/2)\Delta r$ and the number densities $\rho_\beta$. An optional feature controlled by a specific user directive, as explained in Section 3, allows the user to smooth the RDFs prior to printing them, using an established set of simple expressions [19].

## 3. Usage of POLYANA

### 3.1. Input files

Typically, a DL_POLY simulation will need three input files: the CONTROL file where the simulation details are defined, the FIELD file that describes molecular topologies and the force field, and the CONFIG file which contains the initial conditions (coordinates and, optionally, velocities and accelerations). The resultant Molecular Dynamics trajectory is stored in a file called HISTORY. Three out of these four files, namely CONTROL, FIELD and HISTORY, serve as input to the POLYANA post-processing application. If additional input is necessary for the user to exert finer control over POLYANA's computations, this will be inserted at the end of the CONTROL file in the form of special directives, as explained in subsequent paragraphs.

In particular, with every run of POLYANA the CONTROL file will be read. If no POLYANA directives are found below the `finish` DL_POLY directive that marks the end of DL_POLY input, then default values will be used for the bin width, maximum considered distance and other quantities. Otherwise, these values will be defined according to the directives presented in the next section. Then, the FIELD file will be read so that POLYANA knows what molecules are there in the system, their numbers and the interaction sites they are composed of. This information will be used, then, to read the HISTORY file and process the stored trajectory to compute the molecular RDFs.

Some times, the DL_POLY user will have to restart a MD run that has been interrupted for whatever reason, by using the '`restart`' DL_POLY directive. Then, the new HISTORY file will be appended to the old one, but if the user has moved or renamed the old trajectory, the resultant HISTORY file will be lacking a header containing information about the system. POLYANA can handle both kinds of trajectory files, i.e. whether the header is missing or not, without the user having to take any particular action.

**3.2 Usage and description of output**

*3.2.1 Usage without directives*

In order to use POLYANA one will have to run the executable in the directory of the MD trajectory to analyse or in one that is in the system's `PATH`. Then, the program will read the FIELD file to figure out what molecules are there in the system, and it will analyse the trajectory (HISTORY file). With the end of the analysis, two files named RDF (not to confuse with DL_POLY's RDFDAT) and POP, will be created.

In RDF, the centre-of-mass radial distribution functions are given for the various types of molecules, numbered 1, 2, 3, ... according to the order they appear in the FIELD file. E.g, if water and ethanol molecules appear in FIELD in that order, then 1=water, 2=ethanol, and the columns 11, 22 and 12 for the respective $g(r)$ functions will be printed in the RDF file. The default values of 0.1 Å and 12.5 Å will be used for the pair distance bin and maximum distance, respectively – see next Section on how to change these values.

In POP (standing for 'populations'), the number of type $\beta$ molecules around a type $\alpha$ molecule within a given distance, will be printed in a similar arrangement as the columns in RDF. These numbers can be obtained by appropriately integrating the RDFs – actually, $\rho dV g(r)$ – with distance. However, we don't have to carry out the integration; POLYANA will do it for the user.

It is worth noting that POLYANA will process the HISTORY file even if the simulation were incomplete. In such a case a message will be emitted with the end of the computation letting the user know that the trajectory was abnormally terminated – that would not prevent POLYANA from computing and printing some final results in RDF and POP. This is a convenient feature in that it allows users to look at the structure of their systems without waiting for the simulation to be over.

*3.2.2 Usage with directives*

POLYANA can read directives placed at the end of the CONTROL file, after the `finish` DL_POLY directive, to control its execution. A list of all POLYANA directives follows:

| | |
|---|---|
| `polyana` | Marks the beginning of a section containing POLYANA directives |
| `end polyana` | Marks the end of a section of POLYANA directives |
| `start [n]` | Skip configurations 1 to *n*-1 and start processing from *n*-th and beyond |
| `stop  [n]` | Skip (don't process) configurations beyond the n-th |
| `rmax` | Maximum distance for *g*(*r*) calculations |
| `dr` | Distance bin for the histograms in *g*(*r*) calculations |
| `smooth` | If present, *g*(*r*) will be smoothed as discussed above |

The POLYANA directives are case insensitive: `start`, `START` and `Start` are equivalent. Any number of spaces can be inserted before the directives or between the directive keyword and its accompanying numerical value. Their order is irrelevant. The POLYANA and `end` POLYANA lines must exist and enclose the other lines if directives are to be used. If some or all directives are missing, (see: 'Usage without directives', above), default values will be used instead. These are:

| | | |
|---|---|---|
| `start` | = | 1 |
| `stop` | = | [set to Fortran's intrinsic `huge(integer)` to exceed any reasonable number of MD steps] |
| `rmax` | = | 12.5 |
| `dr` | = | 0.1 |
| `smooth` | | [assumes no value; its absence is equivalent to `.FALSE.`] |

*3.2.3 Example of usage with directives*

To further clarify the usage of directives let's consider a concrete example of a MD simulation where 6000 configurations have been stored. The first 1000 configurations belong to the equilibration stage to be excluded from processing. Also, the last 1000 will not be processed. RDFs will be computed for distances up to 10 Å and the bin to be used will be 0.20 Å. Finally, we opt for smoothing the resultant RDF curves. Then, the POLYANA section in the CONTROL file should look like this:

```
...
[various DL_POLY directives]
...
finish [end of DL_POLY section]
```

```
polyana
    start   1001
    stop    5000
    rmax    10.0
    dr       0.2
    smooth
end polyana
```

The indentation does not matter and it is only used here for the purpose of readability.

**3.3 Exploring additional possibilities**

Often times we are interested in intermolecular interactions among certain groups of interaction sites rather than whole molecules. As a concrete example, let's assume we have simulated a system of water mixed with *n*-butanol, using suitable united-atom models. Then, we would like to see how water is arranged around the hydroxyl or the alkyl tail of the alcohol. To do that, we take advantage of the definition of centre of mass, $\sum m_i r_i / \sum m_i$, where *i* runs over the sites in the molecule, to filter the groups of interest by setting all other united-atom masses equal to zero. In the following paragraphs it is assumed that we are using the well-known TraPPE [20] and SPC/E [21] models to model the alkanol and water species, respectively, so the FIELD file will look as follows:

```
...
MOLECULES       2
Butanol
NUMMOLS ...
ATOMS 6
    CH3H        15.0344         0.0000   1
    CH2B        14.0336         0.0000   1
    CH2B        14.0336         0.0000   1
    CH2A        14.0336         0.2650   1
     OC         15.9996        -0.7000   1
     HC          1.0008         0.4350   1
...
...
FINISH

SPCE Water
NUMMOLS ...
ATOMS 3
    OW      15.9996  -0.8476
```

```
    HW           1.0080     0.4238
    HW           1.0080     0.4238
 CONSTRAINTS 3
    1      2    1.0000
    1      3    1.0000
    2      3    1.63298
 FINISH
```

To compute the hydroxyl-water $g(r)$ for the two groups, we rewrite the butanol lines as follows:

```
   CH3H           0.0              0.0000    1
   CH2B           0.0              0.0000    1
   CH2B           0.0              0.0000    1
   CH2A           0.0              0.2650    1
    OC           15.9996          -0.7000    1
    HC            1.0008           0.4350    1
```

Likewise, to calculate $g(r)$'s for the alkyl tail, we simply rewrite:

```
   CH3H          15.0344           0.0000    1
   CH2B          14.0336           0.0000    1
   CH2B          14.0336           0.0000    1
   CH2A          14.0336           0.2650    1
    OC            0.0             -0.7000    1
    HC            0.0              0.4350    1
```

The same trick can be used to 'highlight' a particular site instead of the centre of mass as a basis for the RDF calculation. As another example, geometric centres rather than centres of mass can be looked at by setting masses equal to one and the same value for all site types.

## 4. Example of usage

As an example of practical application of the provided package, we present a test with a system of liquid *n*-hexane in a united-atom representation. The initial configuration, containing 322 *n*-hexane molecules, was obtained from previous simulations of ours and it was subjected to cubic boundary conditions using the TraPPE force field; the initial configuration box length was equal to 40.52 Å. The DL_POLY simulation was carried out at the isothermal-isobaric ensemble, namely at the temperature of $T = 300$ K and atmospheric pressure with a time-step of 1 fs. The calculation lasted 1100 ps, with the first 100 ps being the equilibration stage.

The trajectories were processed with POLYANA using the `smooth` directive and the RDF and POP curves shown in Fig. 1 were obtained. The region between 4 and 8 Å exhibits two overlapping yet visible peaks that probably correspond to different mutual orientations of neighbouring molecules, as sketchily shown in the inset picture of Fig. 1. The 'hack' described in Section 2.3 was then employed to compute intermolecular RDFs for the propyl groups that make up each hexane molecule – that would be useful if one were interested in the development of a coarse-grain force-field mapping three carbon atoms to one 'bead'. The modified FIELD records were as follows:

```
 MOLECULES        2
HexanA
NUMMOLS 161
ATOMS 6
     CH3H        15.0344         0.0000    1
     CH2B        14.0336         0.0000    1
     CH2B        14.0336         0.0000    1
     CH2B        00.0000         0.0000    1
     CH2B        00.0000         0.0000    1
     CH3H        00.0000         0.0000    1
 FINISH
HexanB
NUMMOLS 161
ATOMS 6
     CH3H        00.0000         0.0000    1
     CH2B        00.0000         0.0000    1
     CH2B        00.0000         0.0000    1
     CH2B        14.0336         0.0000    1
     CH2B        14.0336         0.0000    1
     CH3H        15.0344         0.0000    1
 FINISH
```

Since we divide the molecules in two types, above denoted as 'HexanA' and 'HexanB', the output files contain three curves instead of one. These were almost identical; in Figure 2, their average is shown. Normally, one should also 'filter out' the propyl groups the other way around, namely by eliminating the first three united atoms of 'HexanA' and the last three of 'HexanB' and average all results obtained. Because the curves in Fig. 2 are particularly smooth, we trust that the additional RDFs would not differ substantially. The POP curves, on the other hand, should be added. The POP curve in Fig. 2, is actually double the one obtained by POLYANA to compensate for the division of molecules into two groups as above. To interpret the propyl RDF, Fig. 2, it should be borne in mind that POLYANA only computes distances of intermolecular pairs, therefore the propyl connectivity does not show up – this should be helpful when developing a coarse-grain force-field with the appropriate bonded and non-bonded interactions, the latter reflected in the intermolecular distribution functions. Two coordinations shells at around 5 and 9 Å can be clearly seen, with one or two propyl groups in the first and about twelve more in the second one. The system tends to become structureless at distances slightly above the default cutoff value of 12.5 Å used in this test.

An interesting test would be to try and compute united-atom RDFs and compare with the ones obtained by DL_POLY. As said above, POLYANA computes intermolecular distributions whereas DL_POLY will include intra-molecular pairs. Then, the latter will result in peaks characteristic of the molecular connectivity. We carried out this test by computing the $CH_3$-$CH_3$ RDFs. The modified FIELD file was as follows

```
MOLECULES       2
HexanA
NUMMOLS 161
ATOMS 6
    CH3H        15.0344         0.0000      1
    CH2B        00.0000         0.0000      1
    CH2B        00.0000         0.0000      1
    CH2B        00.0000         0.0000      1
    CH2B        00.0000         0.0000      1
    CH3H        00.0000         0.0000      1
FINISH
HexanB
NUMMOLS 161
ATOMS 6
    CH3H        00.0000         0.0000      1
    CH2B        00.0000         0.0000      1
    CH2B        00.0000         0.0000      1
    CH2B        00.0000         0.0000      1
    CH2B        00.0000         0.0000      1
    CH3H        15.0344         0.0000      1
FINISH
```

The obtained curves are compared to the ones computed by DL_POLY in Fig. 3. As expected, additional peaks show up in the DL_POLY distributions, corresponding to the connectivity effect. The sharp peak on the right, at a distance of about 6.5 Å should correspond to the energetically favoured *all-trans* rotational state and the hump at about 5.5 Å should reflect the presence of entropically favoured twisted conformations. These two are completely absent in the POLYANA curve, as the latter only contains the effect of intermolecular pairs. The two RDF curves coincide in the remaining range of pair distances, thus attesting to the reliability of POLYANA output.

## 5. Discussion

In the above, we have presented POLYANA, a new tool for the computation of molecular centre-of-mass radial distribution functions based on trajectories generated by the DL_POLY Molecular Dynamics software. The application has been designed so as to combine ease of use and flexibility in terms of user options, as we have shown in the previous Sections. It is straightforward to use as it does not require any additional input, beyond the DL_POLY input and output files; it can handle abnormally terminated trajectory files so one can obtain results while the simulation is still under way; it allows the user a finer control (time-steps to process, bin width, cutoff distance, smoothing) by means of easy directives inserted in the DL_POLY input files; and the user can make it extract additional information, like intermolecular RDFs of groups of interaction sites or even single sites, by 'hacking' the topology (FIELD) file.

Compilation is a trivial task and, in any case, detailed instructions can be found in the README file that comes with the distribution. GNU make was used by us to compile the program in Linux environments and the included test scripts run in the bash shell; however, it should be easy to build and run the program in other environments and operating systems. The distribution contains some tests from the DL_POLY Classic package to ensure that POLYANA runs correctly; a simple program that was used by the authors to verify the validity of the molecular unfolding algorithm, described in Section 2; and the input files of the liquid *n*-hexane simulation described in Section 4.

POLYANA was initially intended as a companion to DL_POLY. However, handling additional trajectory and topology formats would only require to extend a few particular functions of the input layer. This is indeed our intention concerning next releases; other future goals include: handling more kinds of periodic conditions; implementing appropriate algorithms to extend the RDFs at longer distances [22]; generating tabulated values of inverted-RDF potentials, $-k_\mathrm{B}T \ln g(r)$, and the respective forces (in the form of so-called TABLE files, as regards DL_POLY) as an aid in defining effective coarse-grain pair potentials; adding user input for the definition of groups of sites as an alternative to 'hacking' the FIELD file; computing site-site or group-group inter- and intra-molecular RDFs separately (particularly useful in the case of macromolecular systems); and computing distributions of internal degrees of freedom (bond lengths, bond angles, torsion angles).

Last but not least, the Fortran modules that make up the application require no external dependecies and are so designed that they could be easily read and used in another context, thus serving as a basis for the development of more applications by the users.

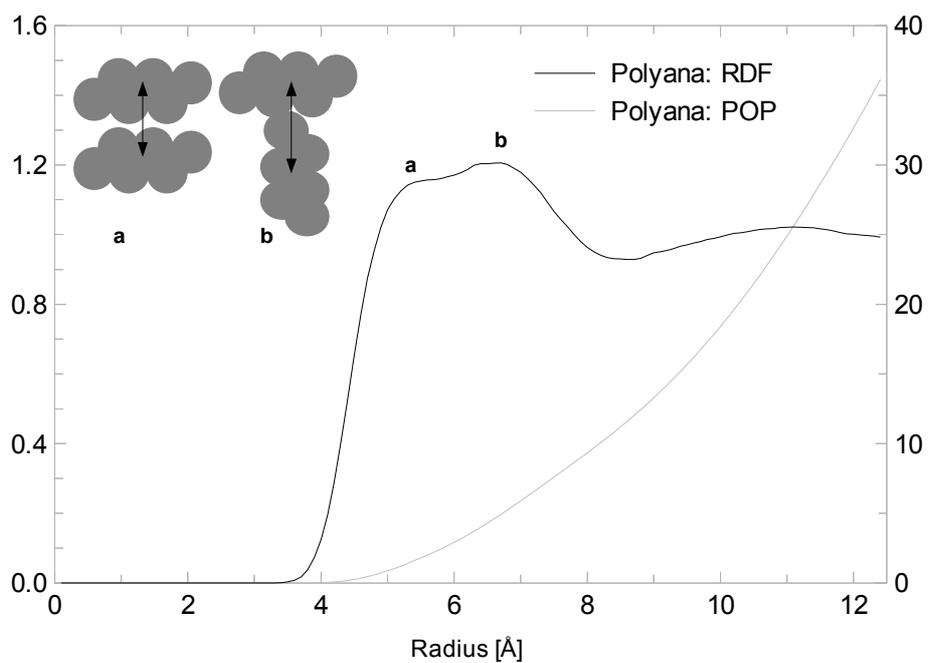

**Figure 1**. Radial distribution function (left-hand side vertical axis) of hexane molecular centres of mass at T=300 K and atmospheric pressure. The average molecular population as a function of distance is also shown (right-hand side vertical axis). Inset picture: two possible mutual molecular orientations corresponding to RDF peaks 'a' and 'b'.

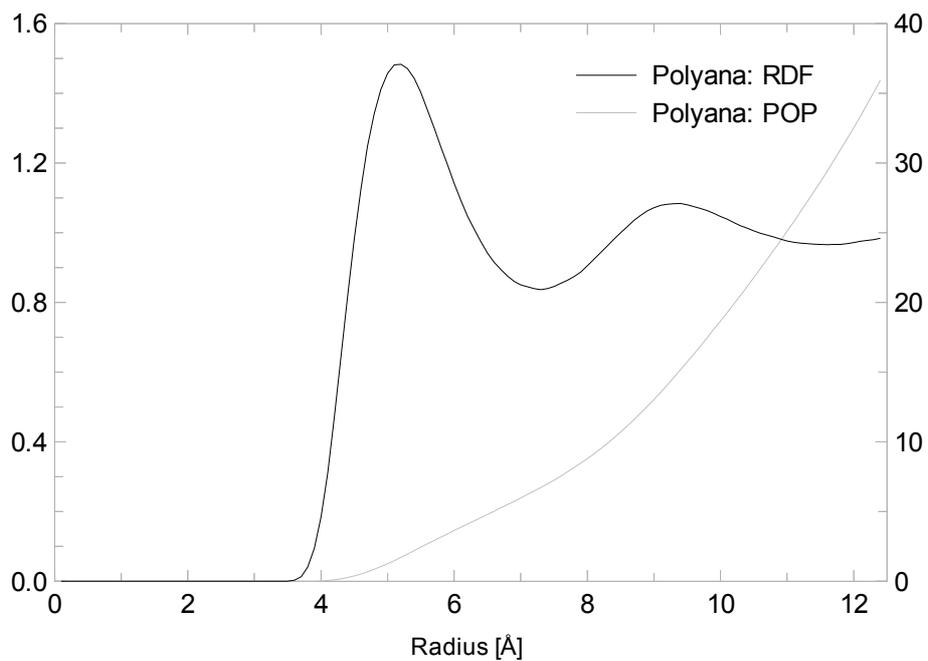

**Figure 2**. Radial distribution function (left-hand side vertical axis) of centres of mass of propyl groups belonging to hexane molecules simulated at T=300 K and atmospheric pressure. The average group population as a function of distance is also shown (right-hand side vertical axis).

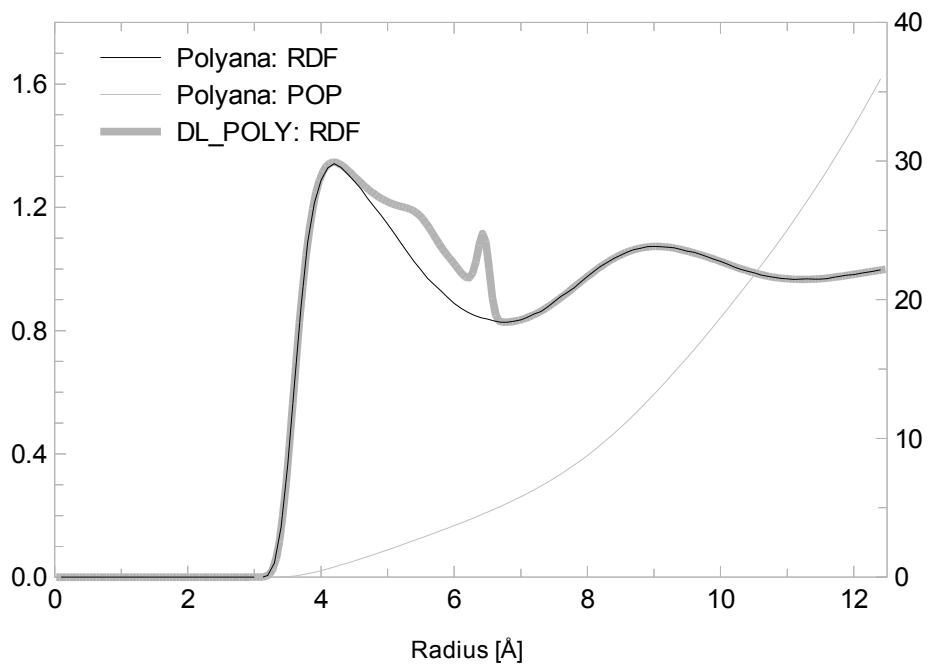

**Figure 3**. Radial distribution function (left-hand side vertical axis) of methyl end-groups belonging to hexane molecules simulated at T=300 K and atmospheric pressure: intermolecular RDF computed by POLYANA as compared to the RDF computed by DL_POLY itself. The average group population computed by POLYANA, is also shown (right-hand side vertical axis).